\documentclass{article}
\usepackage{spconf,amsmath,graphicx,hyperref}
\usepackage{enumitem}
\usepackage{amssymb}
\usepackage{booktabs}
\usepackage{multirow}

\title{Clinical Multi-modal Fusion with Heterogeneous Graph and Disease Correlation Learning for Multi-Disease Prediction}
\name{Yueheng Jiang\textsuperscript{\rm 1}, Peng Zhang\textsuperscript{\rm 1, }\sthanks{Corresponding author.}}
\address{\textsuperscript{\rm 1}Zhejiang University, Hangzhou, China}
\begin{document}
\ninept
\maketitle
\begin{abstract}
 Multi-disease diagnosis using multi-modal data like electronic health records and medical imaging is a critical clinical task. Although existing deep learning methods have achieved initial success in this area, a significant gap persists for their real-world application. This gap arises because they often overlook unavoidable practical challenges, such as modality missingness, noise, temporal asynchrony, and evidentiary inconsistency across modalities for different diseases. To overcome these limitations, we propose HGDC-Fuse, a novel framework that constructs a patient-centric multi-modal heterogeneous graph to robustly integrate asynchronous and incomplete multi-modal data. Moreover, we design a heterogeneous graph learning module to aggregate multi-source information, featuring a disease correlation-guided attention layer that resolves  the modal inconsistency issue by learning disease-specific modality weights based on disease correlations. On the large-scale MIMIC-IV and MIMIC-CXR datasets, HGDC-Fuse significantly outperforms state-of-the-art methods. Our code is released at https://github.com/PhoebeJ9/HGDC-Fuse.
\end{abstract}
\begin{keywords}
Multi-modal fusion, multi-disease prediction, heterogeneous graph, disease correlation learning
\end{keywords}
\section{Introduction}
\label{sec:intro}
Multi-disease diagnosis is a fundamental task in clinical decision-making, where clinicians synthesize complementary evidence from heterogeneous data sources, including electronic health records (EHR) and medical imaging. While recent deep learning methods\cite{joze2020mmtm,polsterl2021combining,hayat2022medfuse,yao2024drfuse} have shown promise in multi-disease prediction by integrating multi-modal data, a persistent gap remains between research models and real-world clinical applicability. This gap is driven by several inevitable challenges that most existing studies fail to fully address:


\textbf{Challenge I: Modality Missingness and Noise.}  Due to clinical or administrative reasons, some modalities are inevitably missing or noisy in practice. For example, imaging such as chest radiographs (CXR) may be absent, and EHR variables can be sparse or intermittently recorded\cite{miller2009privacy}. Some methods attempt to mitigate these issues by synthesizing missing modalities via generative models\cite{ma2021smil,sharma2019missing}, disentangling shared and modality-specific representations\cite{yao2024drfuse,wang2023multi,chen2019robust}, or employing sequence models such as LSTMs to model missingness\cite{hayat2022medfuse}. However, these solutions may result in degraded performance by either amplifying noise or discarding useful signals. Moreover, extending these methods to three or more modalities typically leads to high model complexity and training instability, further limiting their reliability in real clinical settings. 

\textbf{Challenge II: Modality Temporal Asynchrony.} Clinical modalities are collected on different schedules.  In practice, EHR variables are recorded routinely, but imaging like CXR is performed at irregular intervals\cite{al2018daily}. Furthermore, a sequence of CXRs can provide a crucial timeline reflecting disease progression or a patient's response to interventions\cite{borghesi2020covid}. However, current methods\cite{hayat2022medfuse,yao2024drfuse} just pair EHR with the latest CXR. This approach fails to utilize cross-modal temporal dependencies that matter for diagnosis, leading to suboptimal clinical outcomes.

\textbf{Challenge III: Modality Inconsistency in Multi-label Prediction.} Across different target diseases, EHR and CXR may provide discordant evidence. Accurately estimating the contribution of each modality for each disease is essential for resolving conflicts and substantially improving multi-label clinical prediction accuracy. An existing method\cite{yao2024drfuse} applies a modality-level attention with ranking losses, but it overlooks the rich relationship information among labels such as disease co-occurrence and interdependency, which  is of great importance in assisting clinical multi-label diagnosis.

To address these challenges, we propose \textbf{HGDC-Fuse}: Clinical Multi-modal \textbf{Fus}ion with \textbf{H}eterogeneous \textbf{G}raph and \textbf{D}isease \textbf{C}orrelation Learning for Multi-Disease Prediction. First, we construct a patient-centric multi-modal heterogeneous graph designed to maintain robustness under modality missingness. This graph incorporates two distinct edge types: one to model temporal asynchrony by linking cross-modal data with time attributes, and another to enhance representations by connecting similar patients. Next, we propose a type-specific aggregation strategy to preserve the unique semantics of each information source. Building on this, a novel disease correlation-guided attention module explicitly captures label interdependencies to adaptively adjust the importance of each modality for every specific disease. In summary, our main contributions are as follows:

\begin{itemize}[noitemsep,nolistsep,leftmargin=*]
    \item We propose HGDC-Fuse, a multi-modal heterogeneous graph learning framework that addresses modality temporal asynchrony, missingness, and noise. To our knowledge, this is the first work to tackle temporal asynchrony for clinical multi-modal data.
    \item HGDC-Fuse resolves modal inconsistency by leveraging disease correlations to learn disease-specific modality significance.
    \item We conduct experiments on large scale real-world datasets and our results demonstrate the superior performance of HGDC-Fuse over state-of-the-art baselines for multi-disease prediction.
\end{itemize}

\begin{figure*}[!t]
    \centering
    \small
    \includegraphics[width=0.99\textwidth,height=0.34\textheight]{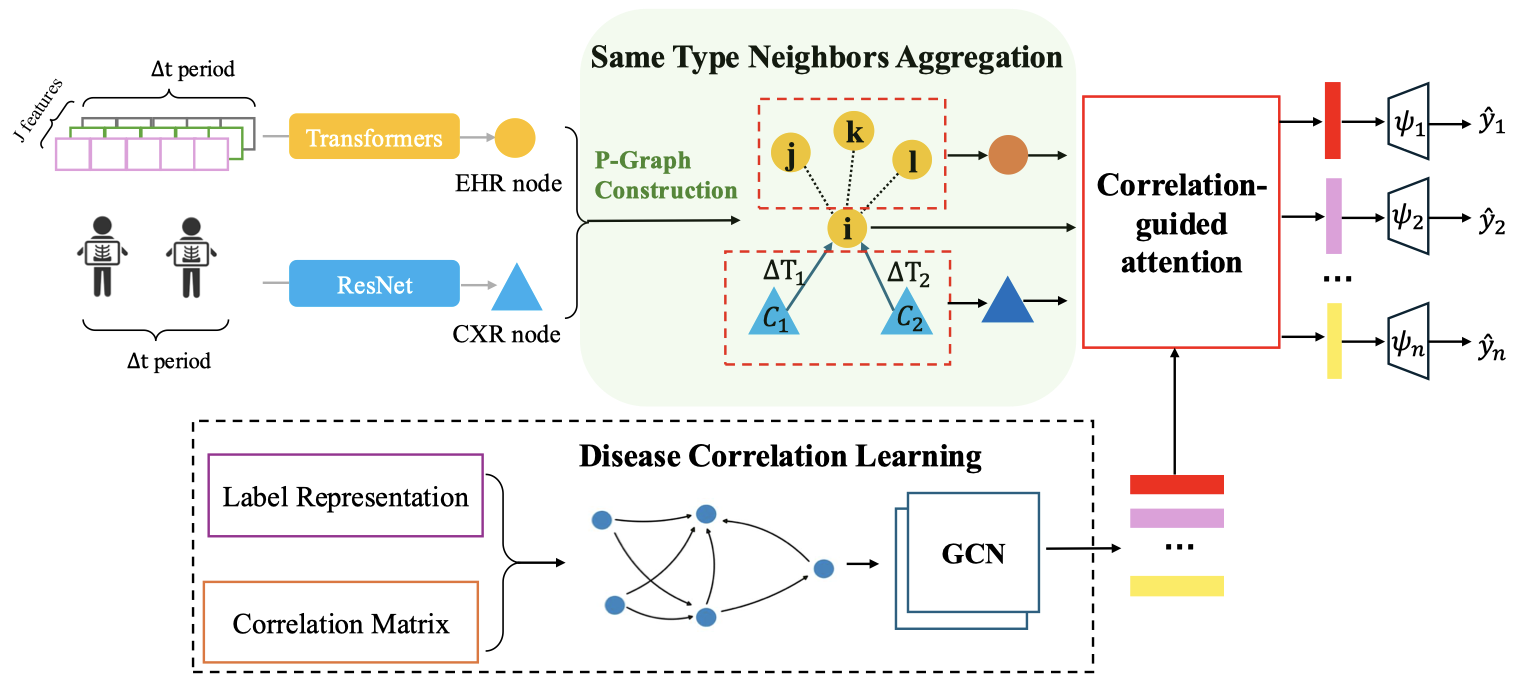}
    \caption{Overall architecture of our model HGDC-Fuse.}
    \label{fig:framework}
\end{figure*}

\section{PRELIMINARY}
\label{sec:format}
In this paper, we consider multi-disease prediction based on two modalities: time-series electronic health records (EHR), and chest X-ray images (CXR). Each patient is associated with a single EHR, while CXR availability varies across patients, ranging from none to multiple images. Let $E_s$$\in \mathbb{R}^{T_n \times J}$ denote the EHR data to patient $s$, $T_n$ and  $J$ are the length of the time series and the number of features, respectively. Let $\mathcal{C}_s=\left\{C_s^1, \ldots, C_s^K\right\}$ denotes the set of all CXR corresponding to patient $s$, where $K$ is the number of CXR.

\section{METHOD}
\label{sec:pagestyle}
The architecture of HGDC-Fuse is presented in Figure~\ref{fig:framework}. In the following subsections, we first elaborate on how to construct a patient-centric multi-modal heterogeneous graph (for addressing Challenge I and II), which effectively captures cross-modal temporal relationships and leverages information from similar patients. We then present three key modules of HGDC-Fuse, i.e., same type neighbors aggregation module, disease correlation learning module, and disease correlation-guided attention fusion module, to aggregate various types of heterogeneous information for each disease based on learned disease co-occurrence  patterns (for addressing Challenge III). Finally, we elaborate on how to make prediction using the fused representation and train HGDC-Fuse.

\subsection{Heterogeneous Patient Graph Structure Construction}
\label{ssec:subhead}
 
We construct a heterogeneous patient graph P-Graph \(\mathcal{G}_s = (\mathcal{V}_s, \mathcal{E}_s)\) for each patient \(s\), where \(\mathcal{V}_s\) is a node set, and \(\mathcal{E}_s\) denotes the edge set. The node set \(\mathcal{V}_s\) contains two types of nodes: EHR node \(n^E_s\) and CXR node \(n^C_s\).  We use a Transformer\cite{vaswani2017attention} encoder to obtain the EHR representation \(\mathbf{h}_s^{ehr}\), and ResNet-50\cite{he2016deep} to extract visual features \(\mathbf{h}_s^{cxr,k}\) from each CXR image \(C_s^k\).

In the P-Graph, the EHR node \(n_s^E\) serves as the target node. Each target node can have two types of neighbors: (1) intra-patient CXR nodes \(n_s^C\), present only when CXR is available, which are connected via directed edges from CXR to EHR with edge weights encoding the relative acquisition times of CXR \(\Delta T(n_s^C)\); and (2) inter-patient EHR neighbors \(n_{s'}^E\), which are constructed by selecting similar patients within the same batch, based on the cosine similarity between their EHR embeddings.


The cross-modal CXR \(\rightarrow\) EHR edges are formally defined as:
\begin{equation}
\mathcal{E}^{cxr \rightarrow ehr}_s = \left\{ (n_i, v_i, \Delta T(n_i)) \mid n_i \in \mathcal{V}_s^{cxr}, v_i = n_s^E \right\}
\end{equation}
And the inter-patient EHR–EHR edges are defined as:
\begin{equation}
\mathcal{E}^{ehr\text{-}ehr}_s = \left\{ (n_s^E, n_{s'}^E) \mid \cos(\mathbf{h}_s^{ehr}, \mathbf{h}_{s'}^{ehr}) > \delta \right\}
\end{equation}
\(\delta\) is the threshold used to determine edge creation.

\subsection{Same Type Neighbors Aggregation}
\label{ssec:subhead}

Within the P-Graph, each EHR node \(n_s^E\) aggregates information from two types of neighbors: other EHR nodes from similar patients, and intra-patient CXR nodes. To preserve the domain-specific semantics and heterogeneous nature of the graph, we devise a type-specific aggregation strategy to obtain two message vectors: \(\mathbf{m}_s^{E \leftarrow E}\) and \(\mathbf{m}_s^{E \leftarrow C}\).

The message \(\mathbf{m}_s^{E \leftarrow E}\) is computed by a multi-head attention mechanism over all neighbor EHR nodes \(\mathcal{N}^E(n_s^E)\):
\begin{equation}
\mathbf{m}_s^{E \leftarrow E} = \bigg\|_{i=1}^H \sum_{n_{s'}^E \in \mathcal{N}^E(s)} \alpha_{ss'}^{(i)} \mathbf{W}_E^{(i)} \mathbf{h}_{s'}^{ehr}
\end{equation}
The attention weights \(\alpha_{ss'}^{(i)}\) are obtained by:
\begin{align}
e_{ss'}^{(i)} &= \text{LeakyReLU}\left(\mathbf{a}^{(i)\top} \left[\mathbf{W}_E^{(i)} \mathbf{h}_s^{ehr} \, \| \, \mathbf{W}_E^{(i)} \mathbf{h}_{s'}^{ehr} \right] \right) \\
\alpha_{ss'}^{(i)} &= \frac{\exp(e_{ss'}^{(i)})}{\sum_{n_{s''}^E \in \mathcal{N}^E(s)} \exp(e_{ss''}^{(i)})}
\end{align}
Here, \(\mathbf{a}^{(i)}\) is a learnable attention vector, and \(\mathbf{W}_E^{(i)}\) is a trainable linear projection matrix for the \(i\)-th attention head.

For intra-patient CXR nodes \(n_s^C\), we design temporal attention weights based on the time embedding of edge \(\mathcal{E}^{cxr \rightarrow ehr}_s\). The normalized time weight for the \(j\)-th CXR node is computed as:
\begin{equation}
w_j^{(s)} = \frac{\exp( \Delta T(n_s^{C,j}))}{\sum_{k=1}^{K} \exp( \Delta T(n_s^{C,k}))}
\end{equation}
Then, the CXR\(\rightarrow\)EHR message is computed as a time-weighted sum:
\begin{equation}
\mathbf{m}_s^{E \leftarrow C} = \sum_{j=1}^{K} w_j^{(s)} \cdot \mathbf{W}_C \mathbf{h}_s^{cxr,j}
\end{equation}
Here, \(\mathbf{W}_C\) is a transformation matrix to be learned.

\subsection{Disease Correlation Learning}
\label{ssec:subhead}

Inspired by the fact that physicians often consider disease co-occurrence patterns in multi-disease diagnoses\cite{barnett2012epidemiology}, we aim to model the complex inter-relationships among diseases to improve clinical prediction. Drawing inspiration from ML-GCN\cite{chen2019multi}, we construct a disease correlation graph where each node corresponds to a disease class.

Each label is represented by a one-hot word embedding vector, forming an initial matrix \(\mathbf{Z} = [\mathbf{z}_1, \dots, \mathbf{z}_N]^\top \in \mathbb{R}^{N \times d}\), where \(N\) is the number of disease labels and \(d\) is the embedding dimension. We compute the correlation matrix \(\mathbf{A} \in \mathbb{R}^{N \times N}\) using conditional co-occurrence statistics extracted from the training set. Specifically, the element \(A_{ij}\) is defined as the conditional probability that label \(j\) occurs given label \(i\):
\begin{equation}
A_{ij} = \frac{\text{co-occur}(i, j)}{\text{count}(i)}, \quad i \neq j
\end{equation}
where \(\text{co-occur}(i, j)\) denotes the number of samples where both labels \(i\) and \(j\) appear, and \(\text{count}(i)\) is the number of samples annotated with label \(i\). To remove noisy correlations, a threshold \(\tau\) is applied:
\begin{equation}
A_{ij} = \begin{cases}
1, & \text{if } A_{ij} >= \tau \\
0, & \text{otherwise}
\end{cases}
\end{equation}

We apply a two-layer Graph Convolutional Network (GCN)\cite{kipf2016semi} to update the label embeddings:
\begin{equation}
\tilde{\mathbf{Z}} = \text{GCN}({\widehat{\boldsymbol{A}}}, \mathbf{Z})
\end{equation}
where $\widehat{\boldsymbol{A}}$ is the normalized version of correlation matrix  \(\mathbf{A} \). The output \(\tilde{\mathbf{Z}} \in \mathbb{R}^{N \times d'}\) serves as disease-aware prototypes that encode higher-order co-occurrence semantics and will later guide disease-specific feature fusion.

\subsection{Disease Correlation-guided Attention Fusion}
\label{ssec:subhead}

We propose a Disease Correlation-guided Attention Fusion module to adaptively fuse multi-source information for each disease. 

Each target node \(n_s^E\) has three types of latent features: 
$\mathbf{h}_s^{\text{ehr}} \in \mathbb{R}^{d}$, 
$\mathbf{m}_s^{E \leftarrow E} \in \mathbb{R}^{d}$, and 
$\mathbf{m}_s^{E \leftarrow C} \in \mathbb{R}^{d}$. These are stacked as:
\begin{equation}
\mathbf{T}_s = [\mathbf{h}_s^{\text{ehr}},\ \mathbf{m}_s^{E \leftarrow E},\ \mathbf{m}_s^{E \leftarrow C}] \in \mathbb{R}^{3 \times d}
\end{equation}

We aim to obtain a label-specific fused representation for each disease $k$ by using the disease label embedding $\mathbf{z}_n \in \mathbb{R}^{d}$ as the query vector. Specifically, we project the features into the key and value spaces:
\begin{equation}
\mathbf{q_n} = \mathbf{W}_q \mathbf{z}_n ,\quad \mathbf{K} = \mathbf{W}_k \mathbf{T}_s,\quad \mathbf{V} = \mathbf{W}_v \mathbf{T}_s
\end{equation}
where $\mathbf{W}_q, \mathbf{W}_k, \mathbf{W}_v \in \mathbb{R}^{d \times d}$ are learnable projection matrices.
The attention weights are computed via scaled dot-product:
\begin{equation}
\boldsymbol{\alpha}_n \;=\; \mathrm{softmax}\!\left(\frac{\mathbf{K}\mathbf{q}_n^\top}{\sqrt{d}} + \mathbf{m}_s \right).
\end{equation}
where mask $\mathbf{m}_s \in \{0,-\infty\}^3$, $m_{s,3}=-\infty$ when CXR is missing. The final disease-specific representation \(\tilde{\mathbf{h}}_n\) is given as:
\begin{equation}
\tilde{\mathbf{h}}_n = \boldsymbol{\alpha}_{n}^{\top} \mathbf{V} \in \mathbb{R}^d.
\end{equation}

\subsection{Prediction and Optimization}
\label{ssec:subhead}
After obtaining the final representations \(\tilde{\mathbf{h}}_n\), the final prediction for the \(n^{\text{th}}\) disease can be obtained using a feedforward layer: \(\hat{y}_n = \psi_n(\tilde{\mathbf{h}}_n)\). We optimize HGDC-Fuse using a Cross-Entropy loss as:
\begin{equation}
\mathcal{L} = \sum_{n=1}^{N} y_n \log(\hat{y}_n) + (1 - y_n) \log(1 - \hat{y}_n).
\end{equation}
where \(N\) is the number of prediction classes.

\begin{table}[b]
\centering
\resizebox{\linewidth}{!}{
\begin{tabular}{lcccc}
\toprule
\multirow{2}{*}{\textbf{Model}} & \multicolumn{2}{c}{\textbf{Trained with the \textit{matched} subset}} & \multicolumn{2}{c}{\textbf{Trained with the \textit{full} dataset}} \\
\cmidrule(lr){2-3} \cmidrule(lr){4-5}
 & \textit{test on matched} & \textit{test on full} & \textit{test on matched} & \textit{test on full} \\
\midrule
\textbf{Transformer} & 0.408 & 0.374 & 0.435 & 0.418 \\
\textbf{MMTM}        & 0.416 & 0.359 & 0.422 & 0.407 \\
\textbf{DAFT}        & 0.417 & 0.348 & 0.430 & 0.409 \\
\textbf{MedFuse}     & 0.427 & 0.329 & 0.434 & 0.405 \\
\textbf{MedFuse-II}  & 0.418 & 0.329 & 0.427 & 0.412 \\
\textbf{DrFuse}      & 0.450 & 0.384 & 0.470 & 0.419 \\
\textbf{HGDC-Fuse}      & \textbf{0.470} & \textbf{0.386} & \textbf{0.489} & \textbf{0.434} \\
\bottomrule
\end{tabular}}
\vspace{0.5em}
\caption{Overall performance measured by the macro average of PRAUC over all 25 disease labels. Numbers in \textbf{bold} indicate the best performance in each column.}
\label{tab:main-results}
\end{table}

\section{Experiments}
\label{sec:typestyle}
\subsection{Datasets and preprocessing}
\label{ssec:Datasets and preprocessing}
We conducted experiments on MIMIC-IV\cite{johnson2019mimic} and MIMIC-CXR\cite{johnson2023mimic}. Following prior work\cite{harutyunyan2019multitask}, our task is to predict 25 disease phenotypes using 17 EHR variables and CXR images from the first 48 hours of an ICU stay. We identified 59,344 stays with EHR, of which 10,630 also had CXRs(avg. 1.89 per stay). We consider two settings: Full (all stays) and Matched (EHR+CXR only). Both are split 7:1:2 for training, validation, and testing.

\begin{table*}[!t]
\centering
\renewcommand{\arraystretch}{1.15}
\resizebox{\textwidth}{!}
{
\begin{tabular}{lcccc}
\toprule
\textbf{Disease Label} & \textbf{CXR (ResNet50)} & \textbf{EHR (Transformer)} & \textbf{DrFuse} & \textbf{HGDC-Fuse} \\
\midrule
Acute and unspecified renal failure & 0.4854 & 0.5129 & \textbf{0.5533 (+7.9\%)} & \textbf{0.5533 (+7.9\%)} \\
Acute cerebrovascular disease & 0.1486 & 0.3976 & \textbf{0.4532 (+14.0\%)} & \textbf{0.4905 (+23.4\%)} \\
Acute myocardial infarction & 0.1610 & 0.1438 & \textbf{0.1756 (+9.1\%)} & \textbf{0.1972 (+22.5\%)} \\
Cardiac dysrhythmias & 0.5777 & 0.4601 & \underline{0.5222 (-9.6\%)} & 0.5966 (+3.3\%) \\
Chronic kidney disease & 0.4191 & 0.4485 & \underline{0.4106 (-8.5\%)} & \textbf{0.4850 (+8.1\%)} \\
Chronic obstructive pulmonary disease and bronchiectasis & 0.3786 & 0.2166 & \underline{0.2709 (-28.4\%)} & \textbf{0.3979 (+5.1\%)} \\
Complications of surgical procedures or medical care & 0.3189 & 0.3679 & \textbf{0.4110 (+11.7\%)} & \textbf{0.4033 (+9.6\%)} \\
Conduction disorders & 0.6116 & 0.1836 & \underline{0.2178 (-64.4\%)} & 0.6390 (+4.5\%) \\
Congestive heart failure; nonhypertensive & 0.6045 & 0.4984 & \underline{0.5420 (-10.3\%)} & \textbf{0.6648 (+10.0\%)} \\
Coronary atherosclerosis and other heart disease & 0.6471 & 0.5617 & \underline{0.5606 (-13.4\%)} & 0.6709 (+3.7\%) \\
Diabetes mellitus with complications & 0.1823 & 0.5054 & 0.5202 (+2.9\%) & 0.5259 (+4.1\%) \\
Diabetes mellitus without complication & 0.2987 & 0.3542 & \textbf{0.3767 (+6.4\%)} & \textbf{0.4006 (+13.1\%)} \\
Disorders of lipid metabolism & 0.5946 & 0.6097 & 0.5862 (-3.9\%) & 0.5974 (-2.0\%) \\
Essential hypertension & 0.5510 & 0.5734 & 0.5790 (+1.0\%) & 0.5983 (+4.3\%) \\
Fluid and electrolyte disorders & 0.5950 & 0.6602 & 0.6638 (+0.5\%) & 0.6867 (+4.0\%) \\
Gastrointestinal hemorrhage & 0.1453 & 0.1069 & \textbf{0.1817 (+25.0\%)} & \textbf{0.2241 (+54.2\%)} \\
Hypertension with complications and secondary hypertension & 0.3992 & 0.4264 & \underline{0.3750 (-12.1\%)} & \textbf{0.4631 (+8.6\%)} \\
Other liver diseases & 0.3601 & 0.2445 & \underline{0.2979 (-17.3\%)} & \textbf{0.4050 (+12.5\%)} \\
Other lower respiratory disease & 0.1759 & 0.1579 & 0.1719 (-2.3\%) & 0.1807 (+2.7\%) \\
Other upper respiratory disease & 0.0998 & 0.1115 & \textbf{0.1346 (+20.7\%)} & \textbf{0.1732 (+55.3\%)} \\
Pleurisy; pneumothorax; pulmonary collapse & 0.1826 & 0.1241 & \underline{0.1698 (-7.0\%)} & \textbf{0.2286 (+25.2\%)} \\
Pneumonia (except that caused by tuberculosis or sexually transmitted disease) & 0.3741 & 0.3671 & \textbf{0.4092 (+9.4\%)} & \textbf{0.4457 (+19.1\%)} \\
Respiratory failure; insufficiency; arrest (adult) & 0.5213 & 0.5855 & 0.5964 (+1.9\%) & 0.6146 (+5.0\%) \\
Septicemia (except in labor) & 0.3974 & 0.5077 & 0.5314 (+4.7\%) & \textbf{0.5464 (+7.6\%)} \\
Shock & 0.3875 & 0.5334 & 0.5524 (+3.6\%) & 0.5570 (+4.4\%) \\
\bottomrule
\end{tabular}
}
\caption{Per-disease PRAUC. The models are trained and tested using the matched subset. The percentages in parentheses indicate the relative difference against the best uni-modal prediction. Differences beyond +5\% are shown in \textbf{bold}, and those beyond -5\% are \underline{underlined}.}
\label{tab:prauc-full-formatted}
\end{table*}

\subsection{Experimental Setup and Baselines}
\label{ssec:subhead}
The model was implemented in Pytorch 2.5.1 and trained on a NVIDIA GeForce RTX 4090 GPU. We set the batch size of full dataset to 256,  the batch size of matched dataset to 64, the similarity threshold \(\delta\) to 0.6, and the correlation threshold \(\tau\) to 0.4. Following \cite{yao2024drfuse}, when training with the matched subset, we randomly remove 30\% of samples that have CXR within each batch as a data augmentation. Due to the highly imbalanced nature of the disease labels, we use Area Under the Precision Recall Curve (PRAUC) to evaluate the performance of HGDC-Fuse and baseline models\cite{saito2015precision}. We compare following baselines: Transformer\cite{vaswani2017attention}, MMTM\cite{joze2020mmtm}, DAFT\cite{polsterl2021combining}, MedFuse\cite{hayat2022medfuse}, MedFuse-II, and DrFuse\cite{yao2024drfuse}. Transformer is a uni-modal method that takes only EHR as input. MedFuse-II is a variant of MedFuse with its CXR encoder and EHR encoder replaced by ResNet50 and Transformer.

\subsection{Overall Performance of Multi-Disease Prediction}

The overall results are shown in Table~\ref{tab:main-results}, where we report macro-PRAUC over 25 disease phenotypes. HGDC-Fuse consistently outperforms all baselines across every setting. Specifically, when trained and evaluated on the matched subset, HGDC-Fuse surpasses DrFuse by 4.4\%, demonstrating HGDC-Fuse's effectiveness in achieving modality fusion. Moreover, when trained on the full dataset and tested on both matched and full subsets, HGDC-Fuse achieves relative improvements of 4.0\% and 3.5\% respectively over DrFuse. These gains highlight the model’s superior robustness under incomplete modality conditions and its ability to effectively leverage partially missing information.

\subsection{Disease-Wise Prediction Performance}
\label{ssec:subhead}
Table~\ref{tab:prauc-full-formatted} presents the disease-wise PRAUC scores for the uni-modal methods, DrFuse, and our HGDC-Fuse, where parenthesized values indicate the relative difference against the best uni-modal prediction. The results highlight that EHR and CXR contribute disparately to the prediction of different diseases. Compared to the uni-modal baselines, DrFuse's performance drops on several diseases. In contrast, HGDC-Fuse consistently outperforms the best uni-modal prediction across nearly all labels. For instance, when predicting \textit{other upper respiratory disease} and \textit{gastrointestinal hemorrhage}, HGDC-Fuse achieves a significant improvement of approximately 55\% for both. These results demonstrate that HGDC-Fuse effectively addresses modal inconsistency by leveraging label correlations to infer disease-specific modal significance, and tackles temporal asynchrony by robustly fusing the two modalities through heterogeneous graph learning.

\subsection{Ablation Study}
\label{ssec:subhead}
To validate each component's contribution in HGDC-Fuse, we conducted an ablation study trained on the matched subset, with results summarized in Table~\ref{tab:ablation study}. We created variants by removing similar patient neighbor nodes in the P-Graph, using only the last available CXR image (discarding temporal information from the image sequence), and replacing the CGA module with a general self-attention method. The performance degradation in the first two variants demonstrates that our multi-modal heterogeneous graphs effectively captures cross-modal temporal relationships and leverages information from similar patients. Furthermore, the superiority of the full model over the w/o CGA variant indicates that the label correlation-guided attention mechanism is crucial, outperforming general self-attention by introducing valuable prior information to resolve modal conflicts.
\begin{table}[ht]
\centering
\small
\begin{tabular}{lcc}
\toprule
Model         & PRAUC             & PRAUC \\
              & @matched subset   & @full dataset \\
\midrule
w/o HER-EHR   & 0.4500            & 0.3812 \\
w/o multi-cxr & 0.4506            & 0.3698 \\
w/o CGA       & 0.4548            & 0.3760 \\
\midrule
HGDC-Fuse     & 0.4698            & 0.3860 \\
\bottomrule
\end{tabular}
\caption{Results of the ablation study}
\label{tab:ablation study}
\end{table}

\section{CONCLUSION}
\label{sec:majhead}
In this paper, we proposed HGDC-Fuse, a novel framework that utilizes heterogeneous graph learning to address critical real-world challenges of modality missingness and temporal asynchrony, while explicitly capturing disease correlations to tackle modality inconsistency in clinical multi-disease prediction. Extensive experiments demonstrate that HGDC-Fuse significantly outperforms state-of-the-art methods. Our work highlights the potential of heterogeneous graphs and disease correlation modeling for developing more robust and reliable multi-modal diagnostic systems for clinical application.


\vfill\pagebreak




\bibliographystyle{IEEEbib}
\bibliography{main}

\end{document}